\documentclass[12pt,showpacs,amsmath,amssymb,pre,superscriptaddress]{revtex4}
\usepackage{epsf}
\usepackage[dvips]{graphicx}

\begin{document}

\title{Clusters die hard: Time-correlated excitation in the Hamiltonian Mean Field model}


\author{Hiroko Koyama\footnote{koyama@gravity.phys.waseda.ac.jp}}

\affiliation{Department of Physics,
Waseda University, Shinjuku, Tokyo, 169-8555, Japan}

\author{Tetsuro Konishi\footnote{tkonishi@r.phys.nagoya-u.ac.jp}}

\affiliation{Department of Physics, 
Nagoya University, Nagoya, 464-8602, Japan}

\author{Stefano Ruffo\footnote{stefano.ruffo@unifi.it}}

\affiliation{Dipartimento di Energetica ``S. Stecco'',
Universit\`a di Firenze, INFN and CSDC, via S. Marta, 3, 50139, Firenze, Italy}

\begin{abstract}

The Hamiltonian Mean Field (HMF) model has a low-energy phase where $N$ particles are
trapped inside a cluster. Here, we investigate some properties of the trapping/untrapping 
mechanism of a single particle into/outside the cluster. Since the single particle
dynamics of the HMF model resembles the one of a simple pendulum, each particle 
can be identified as a high-energy particle (HEP) or a low-energy particle (LEP), 
depending on whether its energy is above or below the separatrix energy.
We then define the trapping ratio as the ratio of the number of LEP to the total 
number of particles and  the ``fully-clustered''  and ``excited'' dynamical states 
as having either no HEP or at least one HEP.
We analytically compute the phase-space average of the trapping ratio by
using the Boltzmann-Gibbs stable stationary solution of the Vlasov equation
associated with the $N \to \infty$ limit of the HMF model.
The same quantity, obtained numerically as a time average, is shown to be in very good
agreement with the analytical calculation.
Another important feature of the dynamical behavior of the system is that 
the dynamical state changes transitionally: the ``fully-clustered'' 
and ``excited'' states appear in turn.
We find that the distribution of the lifetime of the ``fully-clustered''
state obeys a power law. This means that clusters die hard, and that the excitation 
of a particle from the cluster is not a Poisson process and might be 
controlled by some type of collective motion with long memory.
Such behavior should not be specific of the HMF model and appear also in
systems where {\it itinerancy} among different ``quasi-stationary'' states has
been observed. It is also possible that it could mimick the behavior of
transient motion in molecular clusters or some observed deterministic features
of chemical reactions. 

\end{abstract}


\maketitle

\section{introduction}

In systems with long-range interactions~\cite{yellowbook} it is quite common that particle
dynamics leads to the formation of clusters. This happens for instance in self-gravitating
systems \cite{binney}, where massive particles interacting with Newtonian potential, initially
put in a homogeneous state, can create patterns made of many clusters. This phenomenon can be
observed in simplified models, like the one-dimensional self-gravitating systems (sheet models) 
\cite{sheetmodel}. 
For this model an itinerant behavior \cite{itinerancy} between ``quasi-equilibria'' and ``transient''
states has been observed in the long-time evolution~\cite{TGK}. In the ``quasi-equilibrium'' states
particles are clustered, as at equilibrium \cite{rybicki}, but with different energy distributions. 
In the ``transient'' states one particle emitted from the cluster bears the highest energy 
throughout the lifetime 
of the state. The authors of Ref.~\cite{TGK} also claimed that averaging over a sufficiently
long time, which includes many quasi-equilibrium and transient states, should give approximately 
thermal equilibrium.
Motion over several quasi-stationary states is observed also in other  
Hamiltonian systems, like globally coupled symplectic map systems \cite{KK}, or even in realistic
systems of anisotropically interacting molecules \cite{oomine}. 
This shows that thermal equilibrium is not the only possible asymptotic behavior of 
Hamiltonian dynamics. For such cases approaches other than standard statistical mechanics
would be needed. 
Coming back to one-dimensional self-gravitating systems, the generation
of high-energy particles plays an important role in dynamical evolution. However, a difficulty of the model is that the definition of high-energy particle is ambiguous, which is an obstacle to precisely define ``quasi-stationary'' and ``transient'' states.

A time continuous Hamiltonian model for which particle clustering has been studied both from
the statistical and the dynamical point of view is the Hamiltonian Mean
Field Model (HMF) \cite{IK,AR}, which describes the motion of fully
coupled particles on a circle with attractive/repulsive cosine
potential. Recent reviews discussing this model can be found in
Refs. \cite{hmf-review-2002,chavanis}. This model has a second order
phase transition and, in the ordered low energy phase, particles are clustered. However, when the number of particles is finite, some particles can leave the
cluster and acquire a high energy. Hence, the ``fully-clustered'' state has a finite lifetime and
an ``excited'' state appears where at least one particle does not belong to the cluster 
\cite{AR,nakagawa-kaneko-2000-jpsj}.
Therefore, below the critical energy,  we can observe a similar itinerant 
behavior as for one-dimensional self-gravitating systems, between a
``fully-clustered'' state and an ``excited'' state.

In this paper we investigate and characterize the intermittent transitions between these states
during a long-time evolution for the HMF model.
The main advantage of studying this phenomenon for the HMF model
is that the ambiguity to define the dynamical states can be resolved. In fact, 
the equations of motion of each HMF particle can be represented as those of a perturbed pendulum. 
An ordinary simple pendulum shows two types of motion: libration and
rotation. It shows libration when the phase-point is inside the separatrix, and
rotation when it is outside the separatrix.
We then define High-Energy Particles (HEP) of the HMF model as those
particles which are outside the separatrix, and
Low-Energy Particles (LEP) as those which are inside the separatrix. 
This allows us to define a ``trapping ratio'' which takes the value $1$ for the
``fully-clustered'' state and is strictly smaller than $1$ in the 
``excited'' state.
Contrary to an ordinary simple pendulum, the value of the separatrix energy is
not constant in time and hence the trapping ratio can fluctuate in time.
Here, we show that the numerically computed time-averaged trapping ratio
agrees with that obtained by a statistical average performed
for the Boltzmann-Gibbs stable stationary solution of the Vlasov equation
associated to the HMF model \cite{IK,AR}.
However, we find numerically that the probability distribution of the lifetime 
of the ``fully-clustered'' state is not exponential but follows instead a power law.
Therefore, although an average trapping ratio exist, there appear to be no 
typical trapping ratio in the probabilistic sense.

This paper is organized as follows. In Sec.~\ref{sec:model}, we review the
HMF model and define the dynamical states of the system.
In Sec.~\ref{sec:scs} we estimate analytically the trapping ratio, using
a Vlasov equation approach and compare it with the value obtained from numerical simulations.
In Sec.~\ref{sec:lifetime} we numerically compute the probability 
distribution of the lifetime of the ``fully-clustered'' state in order to show that it obeys a
power law. The final section is devoted to summary and discussion.

\section{Model and definition of dynamical states}

\label{sec:model}

In this section we introduce the HMF model and define the dynamical states of the system.
The Hamiltonian of the HMF model \cite{AR} is
\begin{equation}
\label{eq:hamiltonian}
 H=K+V=\sum_{i=1}^{N}\frac{p_i^2}{2}+\frac{\varepsilon}{2N}\sum_{i,j=1}^N[1-\cos(\theta_i-\theta_j)].
\end{equation}
The model describes a system of $N$ particles moving on a circle, each characterized
by an angle $\theta_i$ and possessing momentum $p_i$.
The interaction force between each pair of particles is attractive or repulsive, 
for $\varepsilon>0$ or $\varepsilon<0$, respectively.
In the following we will consider only the attractive case, with $\varepsilon=1$.
In this case, the model displays a second order phase transition at the energy
density $U=H/N=3/4$ from a ``clustered'' phase at low energy (where particles are
clumped) to a ``gas'' phase at high energy (where particles are homogeneously
distributed on the circle). 
The HMF model is a globally coupled pendulum system, and 
the equations of motion can be expressed as those of a perturbed pendulum,
\begin{equation}
\label{pendulum}
 \ddot{\theta}_i=-M\sin(\theta_i-\phi),
\end{equation}
where $M$ (the order parameter of the phase transition) and the phase $\phi$ are defined as
\begin{eqnarray}
\label{eq:m}
 M&\equiv&\sqrt{M_x^2+M_y^2},\nonumber\\
\tan\phi&\equiv&\frac{M_y}{M_x},\nonumber\\
(M_x,M_y)&\equiv&
\frac{1}{N}\left(\sum_{j=1}^{N}\cos\theta_j,\sum_{j=1}^{N}\sin\theta_j\right).
\end{eqnarray}
The single particle energy is 
\begin{equation}
\label{eq:ei}
 e_i=\frac{p_i^2}{2}+[1-M\cos(\theta_i-\phi)].
\end{equation}
Then, the separatrix energy $E_{sep}$ is
\begin{equation}
\label{eq:sep}
 E_{sep}=1+M,
\end{equation}
and the resonance width is $2\sqrt{M}$.

An ordinary simple pendulum shows two types of motion: libration and
rotation. It shows libration when the phase point is inside the separatrix, and
rotation when it is outside the separatrix.
We define High-Energy Particles (HEP) of the HMF model as
those that lie outside the separatrix, i.e. their energy is larger than
the separatrix energy, $e_i>E_{sep}$ . Low-Energy Particles (LEP) lie instead inside
the separatrix, and have energy $e_i<E_{sep}$.
Contrary to the simple pendulum, each particle of HMF model can go from inside
to outside the separatrix and vice versa, because $M$ and $\phi$ are time dependent.

Next, we define the trapping ratio $R$ as
\begin{equation}
R \equiv \frac{N_{LEP}}{N},
\label{def:tp}
\end{equation}
where $N_{LEP}$ is the number of LEP. 
Finally, we define the dynamical states of the system. We say that the system is  ``fully-clustered''
if all the particles are LEP. Otherwise, if at least one particle, among the $N$, is HEP, we say
the system is ``excited''.
The value of $R$ is $1$ if the system is in the ``fully-clustered'' state, it is less
than one if the system is in the ``excited'' state.

\begin{figure}[t]
\includegraphics[width=12cm]{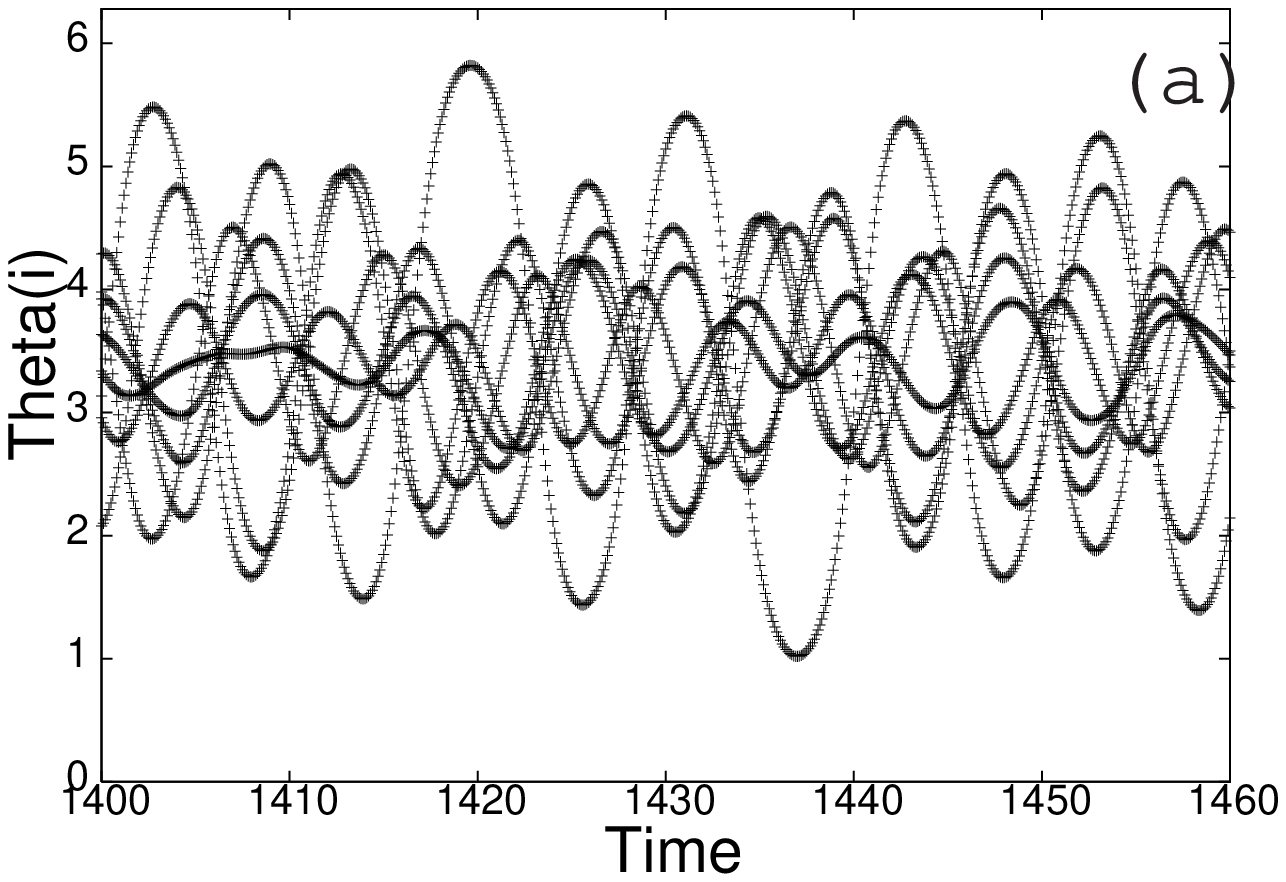}
\includegraphics[width=12cm]{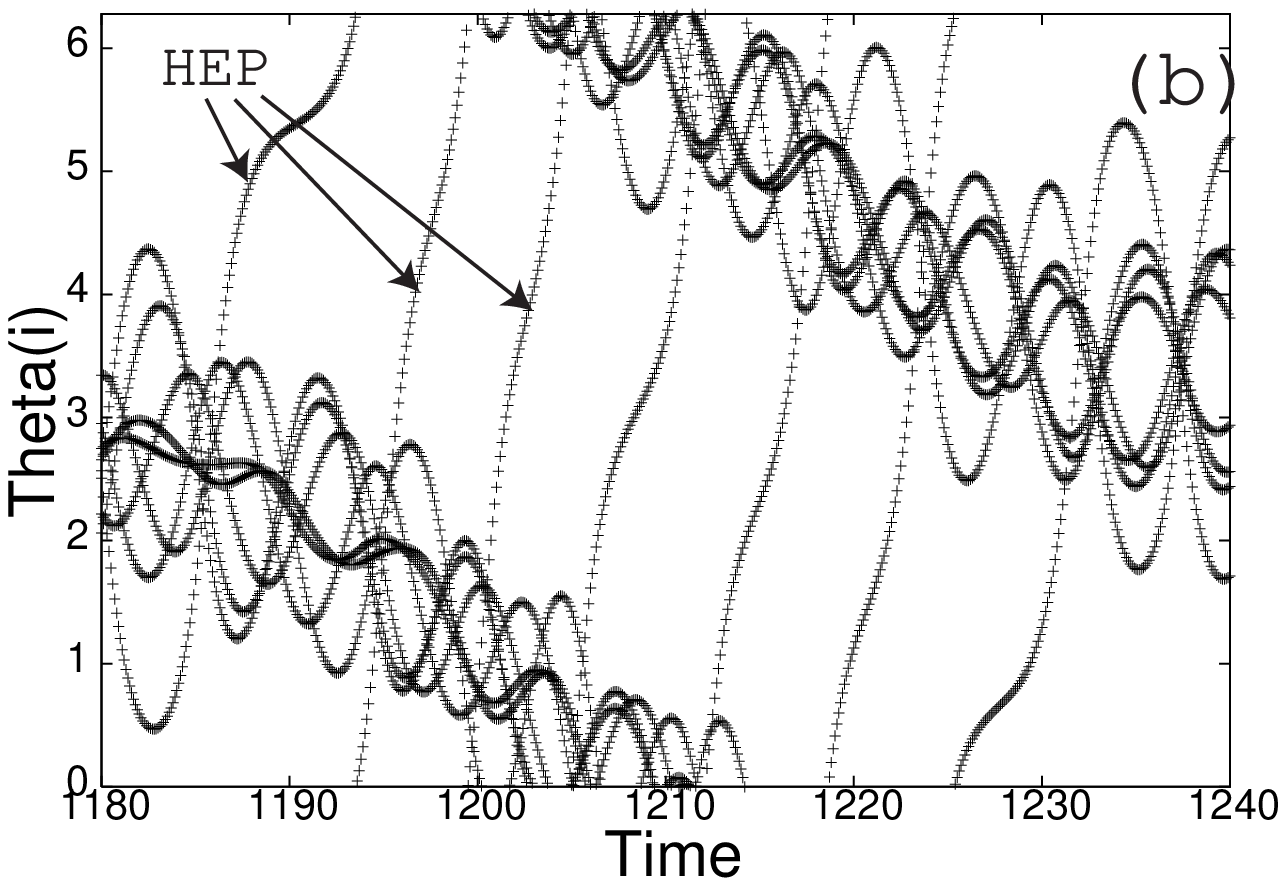}
\caption{a) Time evolution of the positions of the particles
on the circle for the HMF model in a ``fully-clustered'' state; b) same 
time evolution for an ``excited'' state.
Number of particles is $N=8$, system energy density is $U=0.4$ and the initial
condition is a ``water-bag''.  The two states coexist in a given orbit of the system
and appear at different times.}
\label{fig:zahyou}
\end{figure}

Let us discuss an example where these states appear. The system has $N=8$ particles and 
total energy density $U=0.4$ (below the phase transition energy). In the initial condition particles are uniformly
distributed in a square rectangle $[-\theta_0,\theta_0]\times[-p_0,p_0]$
of the single-particle phase-space, with $\theta_0$ and $p_0$
conveniently chosen in order to get the energy $U$. This is the so-called ``water-bag'' initial distribution. As shown in Fig.~\ref{fig:zahyou} the system shows
both the ``fully-clustered'' (panel (a)) and the ``excited'' state (panel (b)) at different time instances, and can switch
from one to the other. In the ``fully-clustered'' state positions of all
the particles fluctuate around a given angle. In the ``excited'' state one HEP has escaped from the
cluster and rotates on the circle (all the others remain clustered). The
momentum extracted by the HEP is compensated by an opposite momentum acquired by the cluster.
This is a typical situation that appears in the low-energy phase of the model. It may be that
more than one particle escapes from the cluster, expecially if the energy is increased.

From this example it is clear that the trapping ratio $R$ is a time fluctuating quantity. In the next
section, we will show that the Boltzmann-Gibbs equilibrium solution of the Vlasov equation associated
to the HMF model allows us to compute analytically the time average of $R$ as a function of the
energy density $U$. The result will be successfully compared with numerical simulations performed at finite $N$.

Another quantity of interest is the time duration of the ``fully-clustered'' state. We define
the lifetime $\tau$ of the ``fully-clustered'' state as the time
interval from the absorption of a HEP to form the full cluster to the
excitation of a particle from the cluster again. We will study numerically the properties of the probability distribution of $\tau$.

\section{Calculation of the average trapping ratio}
\label{sec:scs}
In this section we introduce the Vlasov equation corresponding to the
$N\to\infty$ limit of Hamiltonian (\ref{eq:hamiltonian}) \cite{IK}, and
we derive the stationary stable solution corresponding to
Boltzmann-Gibbs equilibrium. We then show how the knowledge of this solution allows us to derive the average trapping ratio.

The Vlasov equation for the HMF model is
\begin{equation}
\label{eq:Vlasov}
\frac{\partial f}{\partial t}+p\frac{\partial f}{\partial \theta}
-\frac{\partial V}{\partial \theta}\frac{\partial f}{\partial p}=0,
\end{equation}
\begin{equation}
\label{eq:VlasovV}
V(\theta) \equiv \frac{1}{2}
\int_0^{2\pi}\int_{-\infty}^{\infty} \left[1-\cos(\theta-\theta')\right]f(\theta',p')d\theta'dp',
\end{equation}
where $f(\theta,p,t)$ is a single particle distribution function and $V(\theta)$ is the
self-consistent potential.
According to Ref.~\cite{I93}, this Vlasov equation has the following Boltzmann-Gibbs stable stationary solution
\begin{equation}
\label{eq:s-odf}
 f_{BG}(\theta,p)
=\Theta(\theta)\rho(p)=
\sqrt{\frac{\beta}{2\pi}}\exp\left(-\frac{\beta p^2}{2}\right)
\frac{1}{2\pi I_0(\beta \langle M\rangle)}
\exp(\beta \langle M\rangle \cos\theta),
\end{equation}
where $I_0$ is the zero-order modified Bessel function and
$\beta$ is the inverse temperature $\beta =1/T$ ($k_B=1$).
Temperature is twice the kinetic energy $T=2K$.
$\langle M\rangle$ is the solution of the self-consistency equation
\begin{eqnarray}
\label{eq:mbar}
\langle M\rangle
=\left|\frac{I_1( \beta \langle M\rangle)}{I_0( \beta \langle M\rangle)}\right|.
\end{eqnarray}
We easily find that Eq.~(\ref{eq:mbar}) has a non-zero solution only if the temperature 
is sufficiently low, i.e., $T < T_c=1/2$. $T_c=1/2$ (corresponding to $U=3/4$) is the 
temperature of the second order phase transition.

The averaged potential energy is
\begin{eqnarray}
\langle V\rangle&=&\int_0^{2\pi}V(\theta)\Theta(\theta)d\theta
=\frac{1}{2}(1-\langle M\rangle^2),
\end{eqnarray}
and the total energy density is
\begin{eqnarray}
\label{eq:u}
U=\frac{T}{2}+\frac{1}{2}(1-\langle M\rangle^2).
\end{eqnarray}

\begin{figure}[t]
\includegraphics[width=12cm]{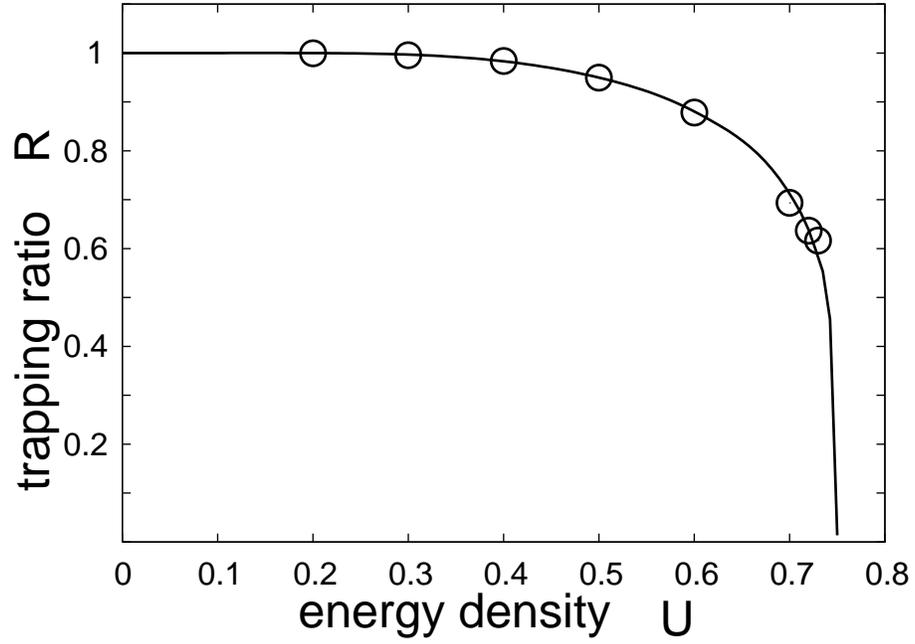}
\caption{Average trapping ratio vs. energy density. The full line is the result
of the analytical calculation of the statistical average $\langle R\rangle(U)$. The points
correspond to the numerical calculation of the time average $\bar{R}(U)$.}
\label{fig:tp}
\end{figure}

In Ref.~\cite{AR}, the energy dependence of the time average trapping ratio
$\bar{R}(U)$ has been numerically calculated. The authors claimed that this quantity 
remains close to unity up to $U=U_b\sim 0.3$ and that it quickly decreases to zero as soon as $U \sim U_c$.
Here, we analytically compute the statistically averaged trapping ratio $\langle R\rangle(U)$, using the 
Boltzmann-Gibbs distribution function (\ref{eq:s-odf}).
The main idea of how to perform this calculation is that of associating the average trapping
ratio to the integral of the single particle distribution function performed inside the phase-space
region $\Omega$ bounded by the upper and lower separatrices of the pendulum motion (\ref{pendulum}). 
\begin{equation}
\langle R\rangle(U)= \int_\Omega f_{BG}(\theta,p)dpd\theta.
\label{idea}
\end{equation}
Using formula (\ref{eq:s-odf}), we obtain
\begin{equation}
\label{eq:tp}
\langle R\rangle(U)=
\frac{1}{2\pi I_0(\beta \langle M\rangle)}
\int_0^{2\pi}{\rm Erf}\left(\sqrt{\langle M\rangle\beta(1+\cos\theta)}\right)
\exp(\beta\langle M\rangle\cos\theta)d\theta,
\end{equation}
where ${\rm Erf}$ is the error function.
The integral in this equation has been performed numerically to obtain the $\langle R\rangle(U)$ 
function plotted in Fig.~\ref{fig:tp}. Numerical values of $\langle R\rangle(U)$ are also reported
in Table~1. The values we obtain for $\langle R\rangle(U)$ are consistent with the time averaged
quantity $\bar{R}(U)$ , first computed numerically in Ref.~\cite{AR}. However, we 
have decided to repeat these numerical calculations for $N=100$ at various energy
densities: the corresponding results are plotted in Fig.~\ref{fig:tp}. 
The agreement between $\langle R\rangle(U)$ and $\bar{R}(U)$ is extremely good.

In order to characterize the finite $N$ fluctuations of the ``fully-clustered''
state, which may create HEP even below the critical temperature, we will analyze
in the next section the probability distribution of the lifetime of the
clustered state.

\section{Power-law distribution of the lifetime of the fully-clustered state}
\label{sec:lifetime}
In this section, we will further characterize the properties of the
fully-clustered state, describing in more detail the trapping-untrapping
process for a small number of particles (a study that has already been
partially done in Ref.~\cite{AR}).

Below the critical energy density, $U<U_c$, HEP are repeatedly excited from 
and absorbed into the cluster in an intermittent fashion. 
The fully-clustered and excited states appear in turn, and the static picture
predicted by the Vlasov equation, where particles that are within the 
separatrices remain there forever, is never observed. Nevertheless, quite
surprisingly, this intermittent state produces, as we have discussed in the
previous section, a time average trapping ratio which is in good agreement
with Vlasov equation predictions.

\begin{figure}[t]
\includegraphics[width=12cm]{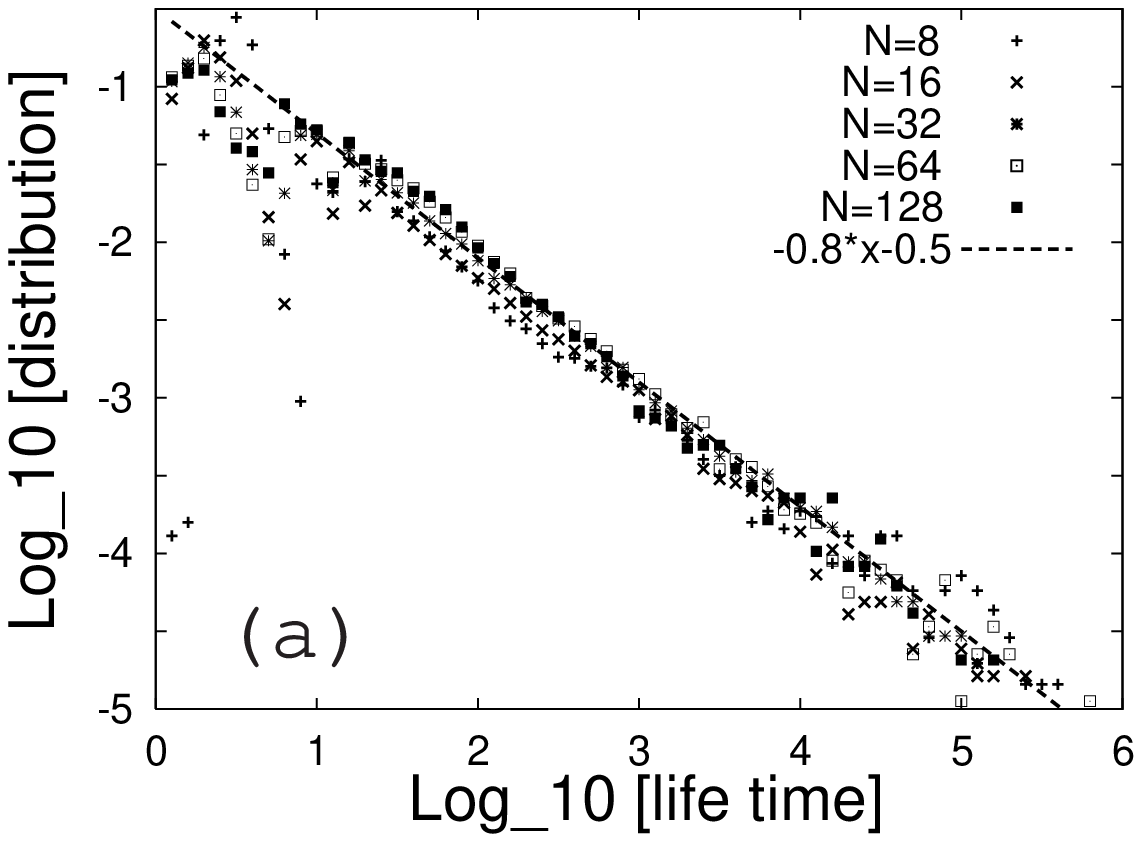}
\includegraphics[width=12cm]{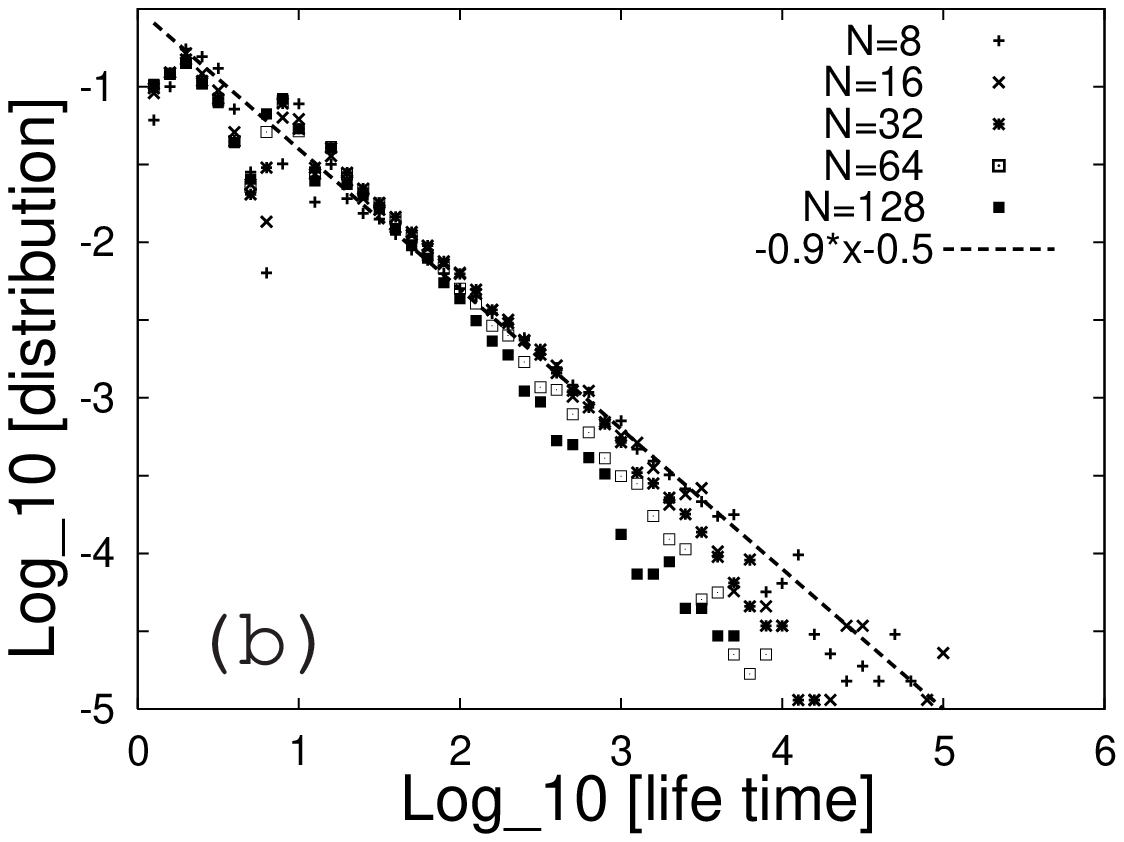}
\caption{The distribution of the lifetime of the fully-clustered state with
total energy $U=0.3$ (panel (a)) and $U=0.4$ (panel (b)). 
Numbers of particles are $8$, $16$, $32$, $64$ and $128$.
The power index is well-fitted by $-0.8$ (panel (a)) and $-0.9$ (panel (b)).}
\label{fig:ltq}
\end{figure}

We want therefore to study the statistical properties of the lifetime
of the cluster, as defined above for a finite number of particles, beginning
with small systems of $N=8$ particles.
If the trapping-untrapping transition process were a
Poisson process, the probability distribution of the lifetime of the 
fully-clustered state would be exponential.
The results of our numerical simulations are shown in Fig.~\ref{fig:ltq}.
The distribution is definitely not exponential but, rather, it obeys a power law,
with an exponent that appears to be slightly dependent on $U$ and to be close to
$-1$.

At fixed $U$, as the number of particle increases, the distribution
is cut-off at large times. This effect is less evident in Fig.~\ref{fig:ltq}a
than in Fig.~\ref{fig:ltq}b. We think that this is due to the difference in
energy between the two cases: at the smaller energy of Fig.~\ref{fig:ltq}a one
would probably need a larger value of $N$ to make the cut-off visible.
Indeed, the presence of such a cut-off time is compatible with the fact that
the average trapping ratio is finite, as shown in the previous section.

In order to get an estimate of the number of particles needed to produce the
cut-off, we propose the following heuristic argument. The single-particle
untrapping probability per unit time can be assumed to be proportional to
$1-\langle R\rangle(U)$. In order to destroy the fully-clustered state, it's enough
that one particle untrap, hence the probability per unit time that the
fully clustered state is destroyed is proportional to $N(1-\langle R\rangle(U))$. When
this probability is of order $1$, the fully-clustered state is destroyed.
A preliminary numerical study of the behavior of $\langle R\rangle$ for
small $U$ gives the non perturbative behavior $\langle R\rangle =1-\exp (-2/U)$.
Therefore, the number
of particle needed to create the cut-off diverges as $\exp (2/U)$ in the limit $U \to 0$,
a growth that is faster than any power. 
The power-law behavior of the lifetime distribution could in principle extend
to infinite time in this limit. This argument could be tested numerically and will
be the subject of future investigations.

\section{summary}
In this paper, we have investigated the dynamical behavior of the Hamiltonian Mean
Field (HMF) model, focusing in particular on the mechanisms of particle trapping
and untrapping from the cluster that is formed in the low-energy phase.
Using the notion of separatrix for the related pendulum
dynamics, we have been able to define precisely the dynamical state of 
``full-clustering'', and the ``excitation'' of a single particle.

We have defined the {\it trapping ratio} as the ratio of the number of particles that
are trapped in the cluster (with energy smaller than the 
separatrix energy) to the total number of particles. 
Using the Boltzmann-Gibbs stable stationary solution of the Vlasov equation,
and performing a phase space integral within the separatrix region, we have
been able to compute analytically the phase-space average of the trapping ratio.
This quantity has been also obtained from numerical simulations of the $N$-body
dynamics of the HMF model and it has been shown to be in perfect agreement 
with the Vlasov equation analytical calculation.

Below the critical energy, when the number of particles is finite, the 
dynamical state of the system changes transitionally: the ``fully-clustered'' and 
``excited'' states appear in turn. That is, high-energy particles (HEP) are excited 
from and absorbed into the cluster intermittently. 
We have numerically computed the probability distribution of the lifetime 
of the ``fully-clustered'' state, finding that it obeys a power law.
This shows that the excitation of a particle below the critical
energy of the HMF model is not a Poisson process.

The discovery of the power law in this system is quite interesting and important.
Its existence implies that, at the moment of ejection of an HEP, the system still 
``remembers'' when the previous HEP had been swallowed into the cluster.
One might think that the system would tend to behave ``thermally'' as the number of
particles is increased, producing an exponential cut-off of the lifetime probability
distribution. Although a cut-off at large times is present in the numerical simulations,
the power law is observed over many decades and its extension increases quite
rapidly with the number of particles. A heuristic calculation based
on the Vlasov equation gives a non-perturbative fast increase of the number of particles
necessary to observe the cut-off.

It will be interesting to investigate the physical origin of
the strong temporal correlations which give rise to the power-law
behavior of the lifetime of the fully-clustered state: a possible collective 
particle motion is a candidate.

Moreover, the power law behavior of the lifetime of clustered states should 
not be specific of the HMF model. It could appear also in systems showing
{\it chaotic itinerancy} among ``quasi-stationary'' states, like those mentioned in the
Introduction \cite{itinerancy} and in models of interacting molecules
\cite{oomine}.

Finally, if we look at the HMF model as representing an ``abstract molecule'', then 
the excitation and decay of HEP could be considered as a chemical reaction.
Then, the power-law type behavior of the lifetime discovered in this paper
implies the possibility of non-standard behavior of chemical reactions:
the reaction rate might be dynamically governed by the coherent motion 
of the reactants~\cite{acp}.

\acknowledgments
H.K. would like to thank Naoteru Gouda for fruitful discussions.
H.K. is supported by a JSPS Fellowship for Young Scientists.
T.K. is supported by a Grant-in-Aid for Scientific Research
from the Ministry of Education, Culture, Sports, Science and Technology.
S.R. thanks JSPS for financial support and the Italian MIUR for funding 
this research under the grant
PRIN05 {\it Dynamics and thermodynamics of systems with long range interactions}.


\begin{table}[h]
\begin{tabular}{|c|c|c|c|}
\hline
$U$&  $M$&$T=1/\beta$&$\langle R\rangle(U)$\\
\hline
0.1  &  0.947209  &  0.0972051  &1.0 \\
0.2  &  0.887109  &  0.186963  &0.999871 \\
0.3  &  0.815506  &  0.26505  &0.996731\\
0.4  &  0.728459  &  0.330652  &0.982999\\
0.5  &  0.621782   &  0.386613 & 0.949673\\
0.6  &  0.485422  &  0.435635  &0.880594\\
0.7  &  0.282056 &  0.479556  &0.712613\\
0.71 &  0.252424 &  0.483718  &0.679443\\
0.72 &  0.21873 &  0.487843  &0.63799\\
0.73 &  0.178692 &  0.491931 & 0.582471 \\
0.74 &  0.126424 &  0.495983 & 0.496192 \\
0.75 &  0.00680319&  0.500046 & 0.0136051 \\
\hline
\end{tabular}
\caption{Statistical equilibrium values of order parameter, temperature and average 
trapping ratio at different energy densities}
\label{tab:num}
\end{table}
\end{document}